\documentclass[twocolumn,showpacs,epsfig,prl]{revtex4}
\usepackage{graphicx}
\usepackage{amssymb}
\usepackage{color}

\newcommand{\figsize}{0.45}

\begin{document}

\draft
\title{Analysis of multiple exceptional points related to three interacting eigenmodes in a non-Hermitian Hamiltonian}

\author{Jung-Wan Ryu$^1$, Soo-Young Lee$^1$, and Sang Wook Kim$^2$}
\email{swkim0412@pusan.ac.kr}
\affiliation{$^1$Department of Physics, Pusan National University, Busan 609-735, South Korea\\
$^2$Department of Physics Education, Pusan National University, Busan 609-735, South Korea}

\date{\today}

\begin{abstract}
We have investigated the exceptional points (EPs) which are degeneracies of a non-Hermitian Hamiltonian, in the case that three modes are interacting with each other. Even though the parametric evolution of the modes cannot be uniquely determined when encircling more than two EPs once, we can recover the initial configuration of the modes by encircling two EPs three times or three EPs twice. We confirm our expectation by numerically calculating the modes of an open quantum system, two dielectric microdisks, and 3$\times$3 matrix model.
\end{abstract}
\maketitle
\narrowtext

Topological properties of quantum mechanics such as Berry phase and Wilczek-Zee holonomy have drawn much attention in physics community \cite{Ber84,Wil84}. Recently a lot of effort has been made to realize quantum computation by using the topologically degenerate ground state of quasi-particles obeying non-Abelian statistics due to its robustness against external perturbations, which is called as topological quantum computation \cite{Kit03,Nay08}. All these topological properties are based upon quantum dynamics described by Hermitian Hamiltonians. However, it has been known that non-Hermitian Hamiltonians also exhibit non-trivial topological nature, which differs from that of Hermitian case in a fundamental level.

Non-Hermitian Hamiltonians have been used to describe open quantum systems in many areas of science. Since every real quantum systems are inevitably open otherwise no information can be acquired from it, usual description using the Hermitian Hamiltonian is only approximately correct. In non-Hermitian case the eigenvalues are complex, and the eigenstates form a non-orthogonal set. The complex eigenvalues have a clear physical meaning; the real and the imaginary part represent the eigenenergy of a state and its decay rate, respectively. Moreover, the degeneracies of the non-Hermitian Hamiltonian, called as exceptional points (EPs), exhibit highly non-trivial characteristics compared with those of the Hermitian one \cite{Kat66,Hei99,Rot09}. When two external parameters are varied so as to form a closed loop enclosing an EP in the parameter space, the two eigenstates, which become degenerate at the corresponding EP, are exchanged with each other. This is topological since it occurs only if the loop encloses the EP irrespective of its precise shape. Such a topological property of EPs has been observed in experiments \cite{Dem01,Lee09,Cho10,Die11}. Recently EPs and non-Hermitian Hamiltonians with PT symmetry have attracted enormous interest \cite{Ben98,Rue10}.

One may ask whether such a topological nature of EPs can be extended into more interacting modes. For three interacting eigenmodes, namely $i$, $j$, and $k$ modes, there exist three possible combinations to form an EP of double degeneracies, namely $EP_{ij}$ for the EP between $i$ and $j$ eigenmodes, $EP_{jk}$ for $j$ and $k$, and $EP_{ki}$ for $k$ and $i$. Here we do not consider the triple degeneracy, which will be briefly discussed later. For an individual EP we expect the topological properties previously mentioned still hold. What then happens if two or three EPs are encircled by one loop? Interestingly this cannot be uniquely determined. If both $EP_{ij}$ and $EP_{jk}$ are encircled, there exist two possible ways to obtain the final result: that is, the exchange between namely $i$ and $j$ first and consequently $j$ and $k$, i.e. $(i,j,k) \rightarrow (k,i,j)$, or the exchange between $j$ and $k$ first and then $i$ and $j$, i.e. $(i,j,k) \rightarrow (j,k,i)$. The former differs from the latter, which naturally raises a question on the non-trivial structure of multiple EPs related to three interacting eigenmodes.

In this paper we show that even though the evolution of the eigenmodes is not uniquely determined as mentioned above, the eigenvalues of the holonomy matrices describing the corresponding evolution reveal peculiar properties so that by encircling two (three) EPs {\em three times} ({\em two times}) one can recover the initial configuration of the eigenmodes. Our theoretical expectation is confirmed by numerically finding quasibound states of an open quantum system, two dielectric microdisks, and $3\times3$ matrix model.

First we briefly explain how the non-trivial topology of the EP appears by using a 2$\times$2 matrix. Consider a non-Hermitian matrix $H$ described as
\begin{equation}
\label{eq:nonHermitianH}
H = \left(
\begin{array}{cc}
E_1 & V \\
W & E_2 \\
\end{array}
\right),
\end{equation}
where $E_1$, $E_2$, $V$, and $W$ are complex numbers. The eigenvalues are given as $E_{\pm}= E_0 \pm \sqrt{\Delta}$, where $E_0=(E_1+E_2)/2$ and $\Delta = (E_1-E_2)^2/4+VW$. The EP takes place when $\Delta =0$ is satisfied, which implies the EP is codimension-2 object. Let us describe small deviation from the EP by a complex number in parameter space, namely $\Delta = \delta e^{i\phi}$, where $\delta (\geq 0)$ and $\phi$ are real. It is noted that in the Hermitian case the degeneracy occurs only if both $E_1=E_2$ and $V=W^*=0$ are fulfilled, implying the degeneracy is codimension-3, and the deviation $\Delta$ should be real and positive. We then continuously vary the external parameters from $\phi$ to $\phi + 2\pi$ so as to encircle the EP in parameter space. The corresponding eigenvalues then rotate around the degenerate eigenvalue, $E_0$ by $\pi$ rather than $2\pi$ in the complex eigenvalue space due to
\begin{equation}
\sqrt{\delta e^{i\phi}} \rightarrow \sqrt{\delta e^{i(\phi+2\pi)}} = e^{i\pi}\sqrt{\delta e^{i\phi}} = -\sqrt{\delta e^{i\phi}}
\label{eq:sqrt}
\end{equation}
implying that two eigenvalues are exchanged; $\{E_0+\sqrt{\Delta},E_0-\sqrt{\Delta}\} \rightarrow \{E_0-\sqrt{\Delta},E_0+\sqrt{\Delta}\}$. Since eigenstates are uniquely determined by a given eigenvalue unless degeneracies occur the corresponding eigenstates have to be exchanged with each other as shown in Fig.~1(b). The so-called holonomy matrix, which is a unitary matrix describing the evolution of eigenmodes after parametric variation along a loop encircling the EP {\em once}, is expressed as
\begin{equation}
\mathbf{M}(1)=
\left( \begin{array}{cc}
0 & 1 \\
1 & 0 \\
\end{array} \right).
\label{eq:EP(1)}
\end{equation}
In order to recover the initial configuration of the eigenstates one should encircle the EP {\em twice} since $\mathbf{M}(1)^2=\mathbf{1}$. In fact, in addition to rearrangement of the eigenmodes the sign changes originating from the geometric phase also occurs \cite{Hei99,Dem01}, for example $M(1)_{12}=-1$ rather than $1$. However, we ignore the sign for simplicity for the time being. We will discuss it later.

\begin{figure}
\begin{center}
\includegraphics[width=\figsize\textwidth]{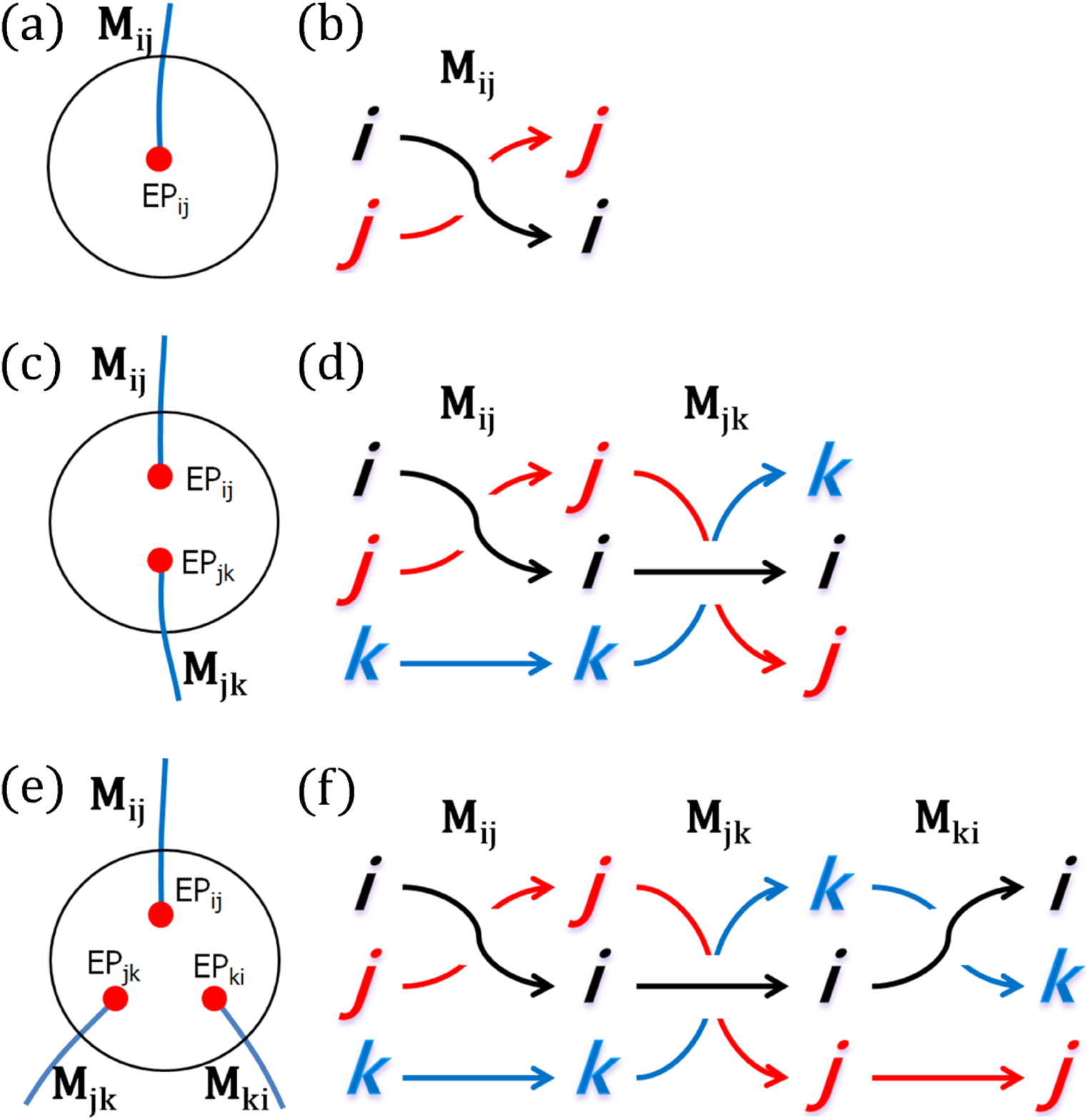}
\caption{(color online) Schematic diagrams describing evolutions of eigenmodes in various situations of encircling EPs. (a) Encircling a single EP in parameter space. Two parameters are continuously varied so as to enclose the loop denoted as the black circle. The blue line attached to the EP represents where the avoided level crossing occurs. The evolution of eigenmode is described by the holonomy matrix $\mathbf{M}(1)$. (b) Two eigenmodes are exchanged by applying $\mathbf{M}(1)$ once. (c) Encircling two EPs. (d) Three eigenmodes are rearranged by successive application of two different $\mathbf{M}(1)$'s, namely $\mathbf{M}_{ij}$ and $\mathbf{M}_{ji}$, resulting in shifting the eigenmodes, that is $(i,j,k) \rightarrow (k,i,j)$. (e) Encircling three EPs. (f) Three eigenmodes are rearranged by applying successive application of three different $\mathbf{M}(1)$'s, resulting in simple exchange of two eigenmodes, that is $(i,j,k) \rightarrow (i,k,j)$.}
\label{pos}
\end{center}
\end{figure}

To analyze three interacting eigenmodes, we need to examine eigenvalues of $3 \times 3$ non-Hermitian matrix, which is completely described by roots of a cubic secular equation, namely $\lambda^3 + a \lambda^2 + b \lambda +c =0$. By using Cardano's method \cite{Kor68}, the general solutions are given as
\begin{equation}
\{\lambda_1,\lambda_2,\lambda_3\} = \{ \alpha_+ + \alpha_- -\beta,\omega \alpha_+ + \bar\omega \alpha_- - \beta,\bar\omega \alpha_+ + \omega \alpha_- - \beta \},
\end{equation}
where $\alpha_\pm = \left( q \pm \sqrt{q^2 + p^3} \right)^{1/3}$, and $\beta = a/3$. Here $p=b/3-a^2/9$, $q=-c/2+ab/6-a^2/27$, and $\omega = (-1+\sqrt{3}i)/2$. Note that $\omega^3=1$. $\bar\omega$ is a complex conjugate of $\omega$. The EP then takes place if $\alpha_+ = \alpha_-$, $\omega \alpha_+ = \alpha_-$, or $\bar\omega \alpha_+ = \alpha_-$ is satisfied, which implies the square root part of $\alpha_\pm$ vanishes. The EPs of three interacting eigenmodes thus has the equivalent topological properties of those of two eigenmodes. It leads us to construct the holonomy matrices of three eigenmodes;
\begin{equation}
\mathbf{M}_{12}=
\left( \begin{array}{ccc}
0 & 1 & 0\\
1 & 0 & 0\\
0 & 0 & 1\\
\end{array} \right)
\mathbf{M}_{23}=
\left( \begin{array}{ccc}
1 & 0 & 0\\
0 & 0 & 1\\
0 & 1 & 0\\
\end{array} \right)
\mathbf{M}_{31}=
\left( \begin{array}{ccc}
0 & 0 & 1\\
0 & 1 & 0\\
1 & 0 & 0\\
\end{array} \right).
\end{equation}
Make sure that $[\mathbf{M}_{ij},\mathbf{M}_{jk}] \neq 0$ as mentioned above.

First let us consider the case of encircling two EPs with a single loop. There exist six possible ways to construct the holonomy matrix denoted as $\mathbf{M}(2)$, i.e. $\mathbf{M}_{12}\mathbf{M}_{23}$, $\mathbf{M}_{23}\mathbf{M}_{12}$, $\mathbf{M}_{23}\mathbf{M}_{31}$, $\mathbf{M}_{31}\mathbf{M}_{23}$, $\mathbf{M}_{31}\mathbf{M}_{12}$, and $\mathbf{M}_{12}\mathbf{M}_{31}$. Interestingly one can obtain the unique secular equation, $\gamma^3=1$, from the eigenvalue problem of all of these holonomy matrices, that is, $\mathbf{M}(2)| \phi \rangle = \gamma | \phi \rangle$, where $|\phi \rangle$ is an eigenstate. It tells us that if one encircles two EPs {\em three} times the configuration of eigenmodes returns to the initial one since $\gamma^3=1$ implies $\mathbf{M}(2)^3 = \mathbf{1}$ no matter what $| \phi \rangle$ is.

As mentioned in the introduction there exist two possible holonomy matrices describing the evolution of eigenmodes when encircling two EPs {\em once} for a given loop because there are two possible orderings of two matrices. To decide which order is correct one should know where the operation performed by $\mathbf{M}_{ij}$ is completed on the path of the evolution in parameter space. Near the EP, roughly speaking, the exchange of two eigenmodes occurs when the avoided level crossing is passed. If we know where the avoided crossing takes place in parameter space, which is represented as a line attached to the EP in Fig.~1(a), (c) and (e), the holonomy matrix can be precisely expected. However, if the loop is far away from the EP, it is not easy to directly apply such an idea since the location of the avoided level crossing cannot be unambiguously identified. Nevertheless, for a given loop encircling two EPs there exist only {\em two} holonomy matrices irrespective of its shape.  It is noted that if the geometric phase is taken into account, for a given loop there exist two holonomy matrices depending on the direction of the rotation of evolution along the loop even in encircling a single EP as will be shown in Eq.~(7) \cite{Hei01}. This also introduces additional non-uniqueness of the holonomy matrices.

Next let us consider the case of encircling three EPs with a single loop, in which there exist six possible holonomy matrices denoted as $\mathbf{M}(3)$, i.e. $\mathbf{M}_{31}\mathbf{M}_{23}\mathbf{M}_{12}$, $\mathbf{M}_{12}\mathbf{M}_{31}\mathbf{M}_{23}$, $\mathbf{M}_{23}\mathbf{M}_{12}\mathbf{M}_{31}$, $\mathbf{M}_{12}\mathbf{M}_{23}\mathbf{M}_{31}$, $\mathbf{M}_{31}\mathbf{M}_{12}\mathbf{M}_{23}$, and $\mathbf{M}_{23}\mathbf{M}_{31}\mathbf{M}_{12}$. Similar to the previous case, $\mathbf{M}(3)$ satisfies the secular equation, $\gamma^2=1$ or $\gamma=1$. In order that the configuration of eigenmodes returns to the initial one, three EPs should be encircled {\em once} for a certain one dimensional solution space or {\em twice} for two dimensional space orthogonal to the previous one dimensional space. This will become clearer in the discussion on 3$\times$3 matrix model later.

We emphasize that our analysis can be directly extended into $n$ interacting eigenmodes, where $n$ is an integer ($n>3$). For example, we consider four interacting modes, in which six EPs exist. If three EPs are enclosed, where their holonomy matrices are written as $\mathbf{M}_{ij}$, $\mathbf{M}_{jk}$, and $\mathbf{M}_{kl}$ using the notation introduced above, one can show that the corresponding secular equation forms $\gamma^4 = 1$. It leads us to the conclusion that if three EPs are enclosed four times, the initial configuration of the modes are recovered.

\begin{figure}
\begin{center}
\includegraphics[width=\figsize\textwidth]{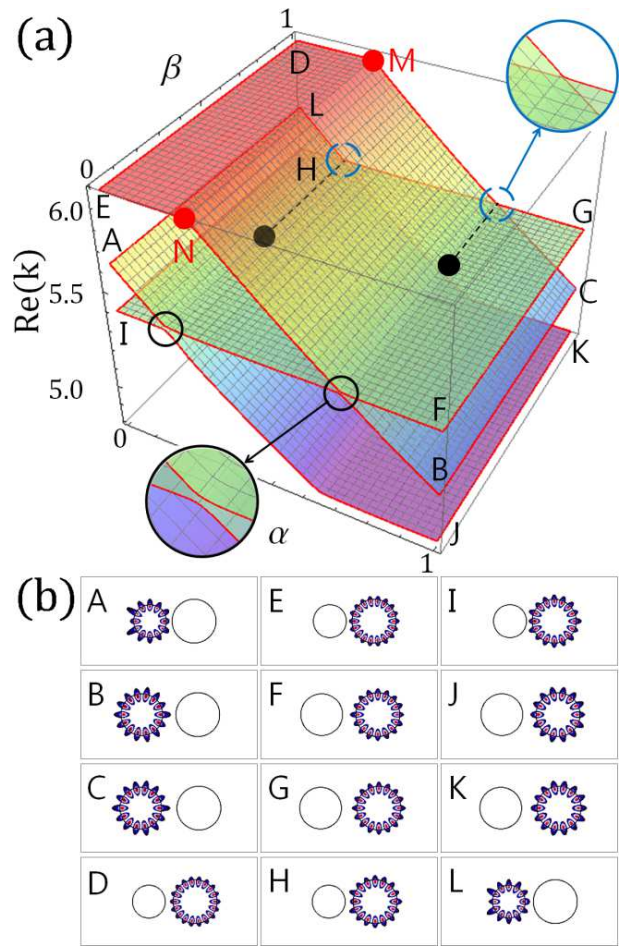}
\caption{(color online) (a) The evolution of real eigenvalues of complex wavenumbers during the parametric variation
along the closed loop enclosing two EPs, namely $\mathrm{EP}_{1}$ and $\mathrm{EP}_{2}$ following the path
$A \rightarrow B \rightarrow ... \rightarrow L \rightarrow A$, which reveals the structure of Riemann surfaces.
The closed loop is formed by a trapezoid with four corners, namely $(0.750, 0.260)$, $(0.950, 0.600)$, $(0.950, 0.700)$, and $(0.750, 0.360)$
in the parameter space $(r, d)$ which corresponds to $(0.2\alpha+0.750, 0.34\alpha+0.1\beta+0.260)$.
Two dots and the attached dashed lines represent EPs and the associated branch cuts, respectively.
The level crossings and the avoided level crossing are pointed out by encircling the dashed and the solid circles, respectively.
The avoided crossings between $W_2$ and ${W_2}'$ modes are denoted as M and N (See the text for detail).
(b) The shape of the eigenmodes for the eigenvalues of A to L designated in (a) in coordinate space.}
\label{surf2}
\end{center}
\end{figure}

In order to confirm our expectation we consider a physical system described by non-Hermitian Hamiltonian. A dielectric microcavity is one of the well-known open quantum systems which have been extensively studied due to its wide range of application \cite{Cha96,Vah03}. The nature of quasi-bound states or eigenmodes of electromagnetic waves confined in the cavities can be intuitively understood by considering the corresponding ray dynamics. The ray inside the cavity can be totally reflected or refracted at the boundary depending on the angle of incidence when the ray hits the boundary. This clearly shows the dielectric microcavity is an open quantum system described by non-Hermitian Hamiltonians \cite{Lee08,Wie08}. Although there exist various shapes of the microcavities, here we consider two closely located microdisks, where two external parameters, the ratio of radii of two microdisks $r$ and the distance of two microdisks $d$, naturally appear \cite{Bor07,Ryu09}. As far as a single microdisk is concerned, the eigenmodes are described by two good quantum numbers, namely the angular quantum number $m$ and the radial quantum number $l$. We are interested in high-$Q$ eigenmodes whose corresponding ray circulates along the perimeter with a large angle of incidence. This is referred to as whispering gallery eigenmode (WGM). Once two microdisks are considered, the WGM of one side starts to interact with that of the other side. However, the eigenmodes of the total system are dominantly localized on one of two disks except near the avoided level crossings \cite{Ryu09}. It leads us to label the eigenmodes as $\mathrm{WGM}_{(m,l)}^{L(R)}$, where $L(R)$ represents that the eigenmode is dominantly localized in the left (right) disk.

Here we deal with four WGM's, $\mathrm{WGM}_{(6,1)}^{L}$, $\mathrm{WGM}_{(7,1)}^{L}$, $\mathrm{WGM}_{(8,1)}^{R}$, and $\mathrm{WGM}_{(9,1)}^{R}$ which are called as $W_1$, $W_2$, $W_3$, and ${W_2}'$. The reason why we consider four rather than three modes is that during the parametric evolution $W_2$ is exchanged with ${W_2}'$, and returned since the loop inevitably crosses the line of avoided crossing of $W_2$ and ${W_2}'$ twice. Therefore, the existence of the mode ${W_2}'$ can be ignored. We numerically find two exceptional points, namely $EP_1$ related to coalescence of $W_1$ and $W_3$, and $EP_2$ of $W_2$ and $W_3$ in the parameter space $(r,d)$. For a given parameter set denoted as $A$ in Fig.~2(a), the distributions of three eigenmodes $W_1$, ${W_2}'$, and $W_3$ in real space are presented in $A$, $E$ and $I$ in Fig.~2(b), respectively. Let us consider the parametric variation along a closed loop encircling two EPs, $EP_1$ and $EP_2$ focusing on the evolution of $W_1$. It is shown in Fig.~2(a) that the real eigenvalue of $W_1$ evolves along the boundary of the surfaces denoted as the red lines during the parametric variation encircling the loop three times starting from $A$. During the first encircling along the loop the eigenvalue follows the path $A \rightarrow B \rightarrow C \rightarrow D \rightarrow E$ as shown in Fig.~2(a) transforming $W_1$ into ${W_2}'$, which is also reproduced by applying $\mathbf{M}(2)$ once. Note that $W_2$ is exchanged with ${W_2}'$ during this evolution as mentioned above. However, ${W_2}'$ will be exchanged with $W_2$ again during the rest of the parametric evolution so that we can effectively ignore ${W_2}'$ for the whole process. Such an exchange between $W_2$ and ${W_2}'$ occurs at M and N in Fig.~2(a). If we encircle the loop once again, the eigenvalue follows $E \rightarrow F \rightarrow G \rightarrow H \rightarrow I$, and ${W_2}'$ is evolved into $W_3$. In order to return to the initial state, $W_1$, we should encircle the loop once again leading to the path $I \rightarrow J \rightarrow K \rightarrow L \rightarrow A$, that is, totally {\em three times}. This is exactly what we expected.

To confirm our expectation for the case of encircling three EPs,
we introduce the $3 \times 3$ matrix $H_3$ described by
\begin{equation}
\label{eq:3by3model}
H_3 = \left(
\begin{array}{ccc}
e_1-i \gamma_1 & \delta & \delta \\
\delta & e_2-i \gamma_2 & \delta \\
\delta & \delta & e_3-i \gamma_3 \\
\end{array}
\right),
\end{equation}
where $e_1$, $e_2$, and $e_3$ are functions of external parameter $\alpha$, while $\gamma_1$, $\gamma_2$, and $\gamma_3$ are functions of $\beta$.
We set $e_1=\alpha-3$, $\gamma_1=\beta-1$, $e_2=-\alpha+1$, $\gamma_2=-\beta+3$, $e_3=0$, $\gamma_3=2$, and $\delta=0.4$, respectively.
Figure~3 shows real parts of three eigenvalues in the ranges of $0.4<\alpha<3.5$ and $1.6<\beta<2.2$.
In the parameter space, there exist three EPs at ($\alpha$, $\beta$)=($1.401$, $1.948$), ($2.072$, $1.686$), and ($2.959$, $2.052$).
If an eigenmode starts from $A$ and continuously evolves along the path $A \rightarrow B \rightarrow C \rightarrow D \rightarrow A$, it finally returns to its initial state after encircling the loop {\em only once}. This exactly corresponds to the case that the eigenvalue of $\mathbf{M}(3)$ is equal to $1$, namely $\gamma =1$. It means that one eigenmode does not change at all by applying $\mathbf{M}(3)$, which is clearly shown in Fig.~1(e) by $i$ eigenmode. On the other hand, if an eigenmode starts from $E$ and continuously moves along the path $E \rightarrow F \rightarrow \cdots \rightarrow L \rightarrow E$, it finally returns to its initial state after encircling the loop {\em twice}. This again corresponds to the case of $\gamma^2 =1$. The similar result is obtained if one starts from $I$.
We note that we can find qualitatively the same results obtained in Fig.~2
if the model of Eq.~(6) is applied to the previous case of encircling two EPs.

\begin{figure}
\begin{center}
\includegraphics[width=\figsize\textwidth]{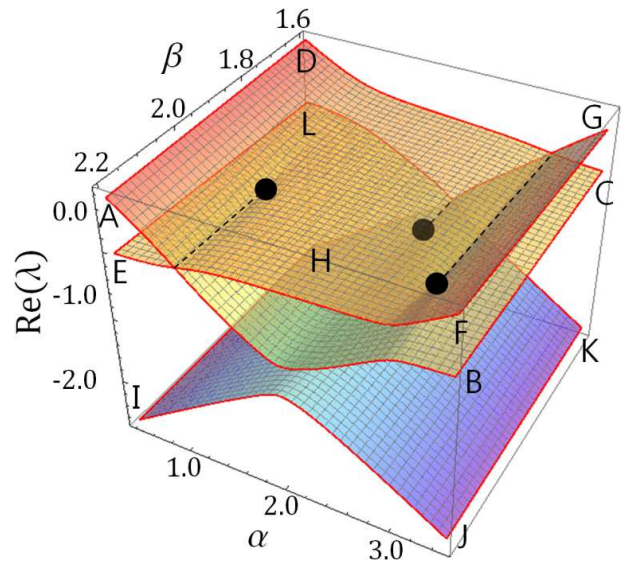}
\caption{(color online) The energy surface of the real parts of the eigenvalues, $\lambda$, of Eq.~(6) (z-axis)
as varying two external parameters ($\alpha$,$\beta$).
Three EPs and the corresponding branch lines are represented as three dots and the attached dashed lines, respectively.}
\end{center}
\end{figure}

For simplicity the geometric phase has been ignored so far. Now let us briefly discuss it. It is known that encircling a single EP generates geometric phase so that Eq.~(\ref{eq:EP(1)}) should be rewritten as
\begin{equation}
\mathbf{M}(1)=
\left( \begin{array}{cc}
0 & -1 \\
1 & 0 \\
\end{array} \right)
{\rm or}
\left( \begin{array}{cc}
0 & 1 \\
-1 & 0 \\
\end{array} \right),
\end{equation}
depending on the rotational direction of the parametric variation along the loop. In this case one should encircle the EP four times to return to the initial configuration due to $\mathbf{M}(1)^2 = -\mathbf{1}$.
For encircling two EPs, however, the geometric phases are all canceled out after cycling the loop three times, implying $\mathbf{M}(2)^3 =\mathbf{1}$ that the sign plays no meaningful role. For encircling three EPs, one obtains $\mathbf{M}(3)^2= \pm \mathbf{1}$.

Final remark is in order. Recently the chiral behaviors of the triple exceptional point (TEP), where three eigenmodes coalesce altogether, has been presented, where encircling TEP shifts three eigenmodes in a cyclic manner like encircling two EPs as shown above \cite{Hei08,Car09}. It is ascribed to the cubic-root singularity appearing in the triple roots of the secular equation of 3$\times$3 matrix. At the first glance, encircling three separate EPs seems to be similar to encircling TEP since in both cases all degeneracies, related to three interacting eigenmodes, are enclosed by a single loop. However, the eigenmode configuration returns to its initial state after going around degeneracies {\em twice} in the case of three separate EPs while {\em three times} in the case of TEP. This is not surprising because TEP is a codimension-5 object. It is unlikely that the 2-dimensional surface including three separate EPs also contains the TEP in 5-dimensional space in general.

In summary, we have investigated the multiple EPs related to three interacting eigenmodes of non-Hermitian Hamiltonians. Encircling a certain EP of two eigenmodes among three has exactly the same properties of a usual EP of two interacting eigenmodes. When encircling more than two EPs once, however, the evolution of the eigenmodes cannot be uniquely defined. Regardless of such an ambiguity, the configuration of the eigenmodes return to their initial one when encircling two EPs three times or three EPs twice. We confirm our expectation for encircling two and three EPs by numerically calculating three interacting eigenmodes of two dielectric microdisks and 3$\times$3 matrix model, respectively. We believe our finding shed some light on study on the non-Hermitian Hamiltonian.

We would like to thank Yeung Seon Lee, Barbara Dietz, and Henning Schomerus for useful discussions. This was supported by the NRF grant funded by the Korea government (MEST) (No.2009-0084606, No.2009-0087261 and No.2010-0024644).

\end{document}